# Machine Learning and Artificial Intelligence in Circular Economy: A Bibliometric Analysis and Systematic Literature Review

**Abdulla All Noman[1,*], Umma Habiba Akter[1], Tahmid Hasan Pranto[1] and AKM Bahalul Haque[2]**

[1]North South University, Dhaka, Bangladesh
abdulla.noman01@northsouth.edu; umma.akter@northsouth.edu; tahmid.pranto@northsouth.edu
[2]LUT University, Lappeenranta, Finland
bahalul.haque@lut.fi
*Correspondence: abdulla.noman01@northsouth.edu

**Abstract:** With unorganized, unplanned and improper use of limited raw materials, an abundant amount of waste is being produced, which is harmful to our environment and ecosystem. While traditional linear production lines fail to address far-reaching issues like waste production and a shorter product life cycle, a prospective concept, namely circular economy (CE), has shown promising prospects to be adopted at industrial and governmental levels. CE aims to complete the product life cycle loop by bringing out the highest values from raw materials in the design phase and later on by reusing, recycling, and remanufacturing. Innovative technologies like artificial intelligence (AI) and machine learning(ML) provide vital assistance in effectively adopting and implementing CE in real-world practices. This study explores the adoption and integration of applied AI techniques in CE. First, we conducted bibliometric analysis on a collection of 104 SCOPUS indexed documents exploring the critical research criteria in AI and CE. Forty papers were picked to conduct a systematic literature review from these documents. The selected documents were further divided into six categories: sustainable development, reverse logistics, waste management, supply chain management, recycle & reuse, and manufacturing development. Comprehensive research insights and trends have been extracted and delineated. Finally, the research gap needing further attention has been identified and the future research directions have also been discussed.

**Keywords:** Artificial Intelligence; Bibliometric Analysis; Circular Economy; Machine Learning; Systematic Literature Review

## 1. Introduction

In a traditional linear economy, which follows the "accumulate-generate-dispose" model, raw materials are administered into necessary products; those products are used for a certain period, eventually converting products into trash. This disruptive production approach has raised serious concerns regarding health, biodiversity, climate, and our overall atmosphere [1]. The linear flow of the current economic structure causes 2$ worth of social, economic and environmental expenditure to

produce 1$ worth of consumable food products[1]. Approximately 60-70% more food will be needed within the year 2050, but the current production structure is unfavorable for soil health and the environment. According to a report by the World Bank, the growth of waste is expected to reach around 3.40 billion tons by the year 2050 [2]. This linearity of the economy and its associated production process is the ruling contributor to waste production [3]. The key solution to these flaws is known as circular economy, which is the idea of diminishing waste production by designing products that can be reused, repaired, refurbished, upgraded and disassembled; hence, extending a product's life cycle to its maximum.

Circular economy efforts to bend the linear progression of gather-manufacture-discard based economy to a circular one where every product is a potential input to another production contemplating products as resources. The idea is to manufacture durable, restorative and reformative products so that goods can be rented, borrowed, or shared wherever practicable. The aim is to trace back the generation of waste to its upstream [4], where waste-generating products are designed and manufactured in the first place. Another cardinal scheme of circular economy is replacing the end-of-life concept with restoration, shifting toward the usage of renewable energy while removing the use of toxic substances that obstruct reuse and return to the biosphere [5]. Thence, circular economy principles can be delineated as i. creating schemes that eradicate pollution, ii. preserving products for as long as practicable and iii. natural arrangements that are reusable [6].

Advanced techniques like artificial intelligence can be used to achieve the circular flow in the economy. Many works on the circular economy have used Artificial Intelligence (AI), Machine Learning (ML), and Data Mining techniques. With the perception of zero-waste production, a nonlinear decision support system has been proposed by Alavi *et al.* [7] where Machine Learning is used to maintain and synthesize the criteria scores for the suppliers in a circular supply chain. Rakhshan *et al.* [8] present a probabilistic prediction model that employs sophisticated supervised machine learning algorithms to assess the reusability of load-bearing building materials after a building reaches its extremity. Artificial neural networks and Nonlinear Regression have been used to build a heat recovery system for the carbon fiber production lineup. Without impairing the mechanical properties, this heat recovery system saves heat that corresponds to as much as 64% of total heat produced, eventually resulting in a circular heat flow along the entire production phase [9]. Artificial intelligence is becoming more prevalent in the fourth industrial revolution period, AI approaches such as machine learning and data mining is being increasingly adopted [10,11].

The volume and area of research related to the circular economy using machine learning approaches have significantly increased, creating a vast amount of bibliometric information. Systematic analysis of this information can provide a comprehensive and organized record exhibiting recent trends, the concentration of research based on geological location, and perceived pioneer personnel of a particular research topic. While both circular economy and machine learning hold profound significance, the amount of systematic review of scientific research is brief. To the best of our knowledge, no systematic review of literature has been conducted on using machine learning techniques to achieve a paradigm shift towards a circular economy. In this research, we presented a bibliometric analysis and systematic study of literature concerning the adoption of machine learning in the circular economy. The contribution of our study is as followings :

- Bibliometric analysis on 104 documents selected via keyword search in SCOPUS database.
- Systematic literature review of selected documents in terms of six criteria: sustainable development, reverse logistics, waste management, supply chain management, recycle-reuse and manufacturing development.
- Finding how AI and ML systems impact circular economy while also uncovering their inter-relation, research trend, research gap and possible future research directions.

The rest of the document is organized as follows. Section 2 discusses the background study. Then section 3 discusses the methodology followed by the bibliometric analysis in section 4 with different subsections. The systematic literature review has been discussed in section 5. The research gap has been investigated in section 6. The Implications have been illustrated in section 7. Finally, we concluded our study by discussing the limitations and future research directions.

---

[1] Cities and circular economy for food, https://emf.thirdlight.com/link/7ztxaa89xl5c-d30so/@/preview/1?o.

## 2. Background Study

### 2.1. Circular Economy

The term Circular Economy was first coined by Turner and Pearce in their study entitled "Sustainable Economic Development," which highlights the interrelationship between different parameters connecting the environment and the economy. Their primary intent was adjusting the traditional economic paradigm, which is based on the practical benefit-cost principle [12]. However, the concept of circular economy has been debated and analyzed by many scholars, practitioners, and policymakers [13]. The fundamental difference between a linear economy and a circular economy is that a product in the linear economy follows the traditional take-make-dispose life cycle, in which raw resources are turned into goods, that end up generating a large amount of waste. In contrast, the circular economy pursues the 3R strategy reduce-reuse-recycle, where the resource to be used is reduced to a minimum, the reuse of products is expanded to a possible maximum. Finally, products are recycled to bring out even more utilities.

Sustainability or sustainable development has also significantly tightened the linkages with the circular economy. These 3R strategy techniques such as reduce, reuse, and recycle are directly aligned with achieving SDGs (Sustainable Developments Goals) declared by the UN. Therefore, a thriving circular economy can benefit every aspect (economic, social, and environmental) of sustainable development. The nature of the circular economy is to limit throughput flow to a level that our environment can tolerate effortlessly [14]. Eliminating carbon emissions while boosting energy efficiency is another significant aspect of this economic structure. The principal objective is economic success, followed by environmental balance keeping; the long-term significance of circular economy is rarely delineated in the literature, resulting in a blurring of their conceptual contours [13,15]. Adopting a circular economy turns limited resources into unlimited recurring resources constrained by natural availability. As previously demonstrated, this economic concept has the significant promise of correcting imbalances in natural resource supply and demand[2]. For this reason, researchers, businesses, and government agencies are all taking an interest in the circular economy.

Current decision-making practices in business are susceptible to ambiguity which often creates risk and makes businesses volatile [16]. Persis *et al.* [16] developed an Ant Colony Optimization (ACO) and Fuzzy ANN-based system, which identified 79 executive business factors contributing to the circular economic impact on the food industry. The factors were later reduced to 39, and their experimentation found the system to be robust and efficient, which mimics a circular economic ecosystem. A framework for identifying technologies that enable circular economy in material-hungry sectors has been developed by Cetin *et al.* [17]. Their framework successfully identified ten digital technologies that regenerate, narrow, slows and close resource loop principle. Load-bearing components that are used to construct the structure of a building can be efficiently reused and recycled by estimating the usage beforehand. Rakhshan *et al.* [8] showed how predictive supervised ML systems could bring about a circular economic arrangement in construction management, one of the most waste-producing industries awaiting reform. Ramchandani *et al.* [18] devised an incentive mechanism based on AI and Blockchain to tackle the packaging wastage created by large multinational enterprises. The data-driven approach [19] has also been shown by researchers where social, political and economic factors of influence have been studied.

### 2.2. Artificial Intelligence

An intelligent economic system continuously demands innovative solutions to boost the quality and sustainability of actions relate to the whole system while diminishing costs. In this circum- stance, technologies driven by artificial intelligence are ready to render in the new industrial paradigms [20]. John McCarthy [21] is called the father of artificial intelligence. In 1990, he characterized artificial intelligence as follows:

> "Artificial intelligence is the science and engineering of making intelligent machines, especially intelligent computer programs."

---

His speculation of artificial intelligence (AI) establishes AI as a set of functionalities of a computer or a computer-managed automation system which executes tasks that intelligent beings might commonly accomplish [21]. Artificial intelligence is organized into various subsets like data mining, machine learning and deep learning. Artificial intelligence is a combination of mathematical reasoning and error-reducing functionalities [26]. AI techniques primarily learn from a variety of data in the form of numeric entries, audio, video or image and then reduces the error from it learning function. The aim is to make the model as accurate as possible.

In recent times, the area of artificial intelligence has also been blended with several fields such as engineering, economics, education, science, business, medicine, marketing, finance, accounting, law, and the stock market [20,22]. The coverage of artificial intelligence has expanded immensely since the machines have become more computationally powerful over the last decade. The cognitive abilities of different machine learning algorithms, has generated profound impressions on the government, business, and to the society [22]. The contribution of AI has been proven significant towards making the ground to achieve global sustainable development [24,28,48]. Therefore, artificial intelligence can be a great aid in solving vital issues related to an intelligent circular economic system (e.g., sustainable manufacturing system [9], waste management [23,24], reverse logistics [29], optimization of energy sources [79], supply chain management [26]). In recent years, there has also been an ongoing trend to combine artificial intelligence with all aspects of CE [9,23,24].

### 2.3. How AI and ML Aids Circular Economy

Artificial intelligence (AI) is a term used to describe intelligent systems that attempt to gain human-like cognitive skills such as the capacity to reason, understand the meaning, generalize, and understand from previous incidents. Artificial intelligence (machine philosophy of thinking, acting, and performing) has drawn the attention of major industries of the current economy. On the other hand, Circular Economy (CE) has gained traction among academics, governments, and business communities. The sustainability of an organization is now firmly tied to digital innovation. The central concept of the circular economy stands on digitization and innovation that help secure sustainability in the long run. Digital innovation facilitates a circular economy that ensures maximum use of limited resources using digital platforms, smart devices, and artificial intelligence [25].

AI is one of the paramount technologies which can provide numerous advantages with the assistance of various algorithms for a smooth transition towards CE. For instance, real-time data analysis[26] for supply chain management, cost reduction[27] and carbon footprint reduction[28] for sustainable development, automate the processes[29] for reverse logistics, assessing the impact of waste generation[30,31] for waste management, sorting different materials[23] for recycling purposes etc. Moreover, AI also has the potential to be an essential enabler of systemic transformation, such as the transition from raw data to decision-making strategies. Many research has been published addressing these issues, where researchers focused on converting the linear economy model to a CE model by embracing different strategies related to recycling, reusing, and remanufacturing [32-34]. In CE, AI can also enable autonomous and remote monitoring of manufacturing efficiency and product end-of-life cycles [35]. According to Ramadoss *et al.* [36], a substantial amount of data originates during the manufacturing, during usage, and during disposal process of a product, component, or material. AI can be utilized to effectively analyze this data for further development of these processes. Ghoreishi *et al.* [35] found that circular design tools and methods can help businesses improve product circularity according to artificial intelligence. Supply chains, logistics, and asset management technologies have also seen major changes with the increasing use of AI, IoT concepts, 3D printing, Advanced Robotics, wearable devices, and Augmented Reality (AR)[36].

### 3. Methodology

This study comprises qualitative and quantitative analysis in the form of bibliometric analysis followed by a systematic literature review. For the bibliometric analysis, we conducted a literature search in the Scopus database in October 2021. The Keywords we applied to find the literature were as follows:
"Artificial Intelligence AND Circular Economy"

OR

"Machine Learning AND Circular Economy"

The initial search resulted in 136 documents matching the search keyword. Among these 136 documents, we employed exclusion criteria to weed out a few publications that did not fit our study objectives. Only research articles, conference papers, and book chapters that match our research objective were selected for further experimentation. Documents such as editorial, retracted, and books (as book compiles various chapters) were discarded as they do not align with our experimentation objective. In addition, we also removed review articles and conference reviews as our work itself is a comprehensive review study. Documents written in languages other than English (i.e., two Chinese documents were found) were also excluded from the selected list. After applying these exclusion criteria, 104 papers remained for the bibliometric analysis.

For the systematic literature review(SLR), we vividly explored the 104 documents that were finally selected for the bibliometric analysis. These 104 documents were further filtered, and only those studies were chosen in which AI has been applied in a practical scenario to solve the barriers related to the circular economy. After skimming the titles and abstracts, we found that 53 papers were irrelevant and out of scope for the SLR. Therefore, the remaining 51 papers were selected for conducting the SLR. We rigorously went through these 51 papers by reading the full text. Finally, 40 documents matched our criteria for this SLR. Figure 1 represents our searching criteria, outcome, selection and exclusion process for the bibliometric analysis followed by SLR.

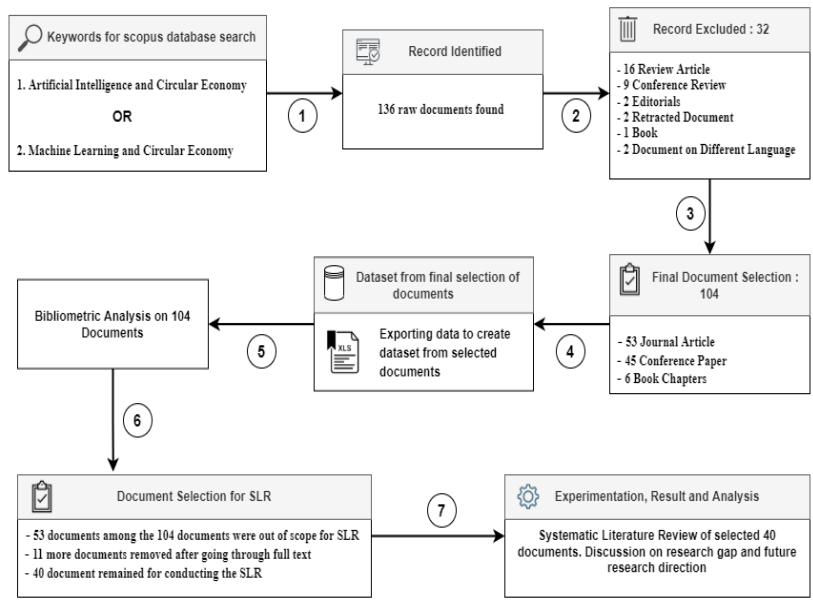

**Figure 1.** Document searching, selecting and exclusion process for bibliometric analysis followed by SLR

## 4. Bibliometric Analysis

The major concern of this study is to find how the integration of AI and ML techniques is aiding different industries to gain CE. As the topic is relatively new and an incremental research trend has appeared in the recent past, the need for an organized bibliometric analysis has been awaited. The process of bibliometric analysis, also known as "scientometrics," is a popular scientific method. It is a quantitative analysis process to observe the relationship of different parameters on a specific area or field [37,38] of research. This method is also used to discuss or visualize the hike of a particular research domain at a specific time [39,40]. SMACOF (Scaling by Majorizing a Complicated Function) is the algorithm used to extract this prevalent information from collected documents. The algorithm first carries out an optimization strategy from multidimensional data as follows.

$$\sigma(x) = \sum_i \sum_j \omega_{ij} \; (\delta_{ij} - d_{ij}(x))^2 \; \dots\dots\dots\dots\dots\dots\dots\dots\dots\dots\dots\dots\dots\dots\dots\dots\dots\dots\dots\dots\dots\dots\dots\dots\dots\dots\dots\dots\dots\dots \; (1)$$

In the equation above (Equation 1), $\sigma(x)$ is a distance measuring loss function that maps the distance between two points from a multidimensional data space X. A weight matrix $\omega_{ij}$ multiplied by the difference between the Euclidean distance $d_{ij}$ and ideal distance $\delta_{ij}$ squared yields the loss function $\sigma(x)$ from n-dimensional data space [41]. An iterative approach of this loss function suggested by De Leeuw [41] is called the SMACOF algorithm, which is popularly used for bibliometric analysis. In this study, we collected data from the SCOPUS database and arranged it into four sections to focus on various aspects and see their relationship. These four sections are primary information, country, author and subject area. These sections were also further divided into various sub-categories,   explained in section 4. Crucial information such as surge research topic, author contribution, the country where particular research topic is booming was found by our bibliometric analysis using these documents. We used an open-source application called VOS viewer to conduct the bibliometric analysis for experimentation. The experimentation details and findings are thoroughly presented as follows.

### 4.1. Primary Information of the Documents

As we previously specified in the methodology section, our work consists of 104 documents. These documents were published in 70 publication mediums like journals and conferences, where 1184 total keywords comprised of author keyword (found in the keyword section of a document) and index keyword (search engine optimization keyword) were found. The number of author's keywords was 384, and index keywords were 932 detected by VOS Viewer. A total of 384 authors participated in these 104 studies, where the average number of authors per document is 3.25. It was found that 17 documents are single-authored while the rest are multi-authored. A total of 39 countries' documents have been found where collaboration between 32 countries has been seen. The number of organizations that are engaged in this research is 230. The number of citations for these 104 documents is 622, with an average citation of 5.98 per document. Prominent research done by D'Amato *et al.* [57] alone got 244 citations, and the rest of the 98 documents got 378 citations. Total cited references for these 104 documents were 4878, totally cited published sources were 2565, and total cited authors were 9434. The primary information accumulated from the 104 documents has been delineated in table 1.

### 4.1.1. Types of Documents and Sources

This section illustrates the information about the types of documents we used and the sources we discovered in our study. From a total of 104 documents, 53 journal articles, noted to be the highest among others, followed by conference papers that count to several 48 documents, and the rest of the six documents are book chapters. We desirably excluded the review papers from our documents as our work itself is a review paper. We also excluded the rest of the types of documents from being very specific in this field. Figure 1 also depicts that the researchers were almost equally interested in writing journal articles and conference papers.

**Table 1.** Primary Information

| Primary Information Category | Total Count |
|---|---|
| Total document | 104 |
| Total Sources | 70 |
| Total Number of Keywords | 1184 |
| Author Keywords | 384 |
| Index Keywords | 932 |
| Total Authors | 338 |
| Single Authored Documents | 17 |
| Multi Authored Documents | 87 |
| Total Organizations | 230 |
| Total Countries | 39 |
| Total Collaborated Countries | 10 |
| Total citations of the documents | 622 |
| Total Cited References | 4878 |
| Total Cited Sources | 2565 |
| Total Cited Authors | 9434 |
| Average Authors per Documents | 3.25 |
| Average Citations per Documents | 5.98 |

Other than this, table 2 depicts the top 15 sources of published documents. "The Journal of Cleaner Production" (UK) and "Sustainability" (Switzerland) were found to publish the most significant number of documents till now. Both of them published a total of six journal articles." AIM- SEC, 2011 (2011 2nd international conference on artificial intelligence, management science, and electronic commerce)" published five documents and placed second." IFIP AICT (Advances in Information and Communication Technology)" and" Procedia CIRP" are in the third position with four published documents each." Lecture Notes in Computer Science" and" Technological Forecasting and Social Change" appeared in the fourth place, having three documents published in this field. The other sources mentioned in table 2 published at least two documents. Therefore, researchers worldwide working with the subject matter can target one of these sources to publish their documents in the future.

**Table 2.** Top 15 Publication Sources

| Name of the Sources | Number of Documents |
|---|---|
| Journal of Cleaner Production Sustainability (Switzerland) | 6 |
| Proceedings of the 2nd International Conference on Artificial Intelligence Management Science and Electronic Commerce (AIMSEC 2011) | 5 |
| IFIP Advances in Information and Communication Technology<br>Procedia CIRP | 4 |
| Lecture Notes in Computer Science<br>Technological Forecasting and Social Change | 3 |
| ACM International Conference Proceeding Series Advances in Intelligent Systems and Computing Clean Technologies and Environmental Policy Computers and Chemical Engineering<br>Energies<br>International Journal of Production Research Journal of Environmental Management Journal of Physics: Conference Series | 2 |

### 4.1.2. Publication Per Year

The total number of publications issued per year can be an inventory of research trends in a designated research field. The summary can also give an overview of the trends in that field in the upcoming future and, apart from that, also shows the growth of a research topic within a specified time frame. Our study observes that the field of" integrated AI and ML with the circular economy" is comparatively new. The first paper of this field was published in 2009, "Research on key technologies of knowledge-based engineering decision support systems for a circular economy." In this paper, the authors proposed a software system that can dynamically monitor the conditions of regional sustainable development [42]. After that, we see a minimum growth of development in this field where the year 2011 seemed to be the beginning. However, surprisingly, in 2013 and 2014, there was no single paper published in this field,  but from 2015 to 2018, there was slight movement from the researchers worldwide. From the year 2019, we see notable growth in this domain. From figure 2, it can be seen that the cumulative publication was increasing rapidly from this year (2019). The maximum number of publications has been seen in the year 2021, and at the time of conducting this study, 2021 has passed. Obviously, the publication number is expected to surpass the previous amounts. As the cumulative publication has been seen increasing in recent years, it is a clear indication that researchers worldwide are interested in this field, and we are most certainly expecting to see many improvements over the previous study in this specific area; hence, producing more documents.

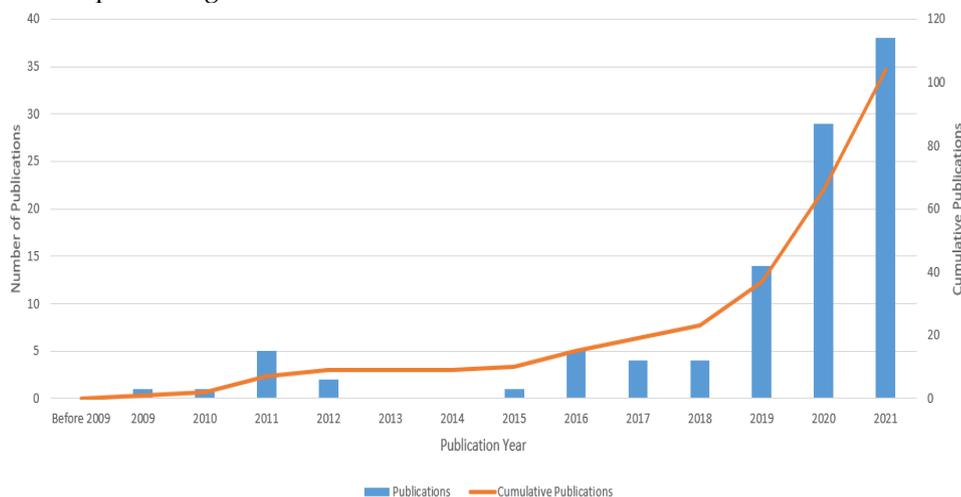

**Figure 2.** Publication per Year in this Domain

### 4.1.3. Citation Per Year

The number of citations per year shows the same trend as document per year, with a slight difference. Figure 3 depicts the yearly citation of the publications. It is highly expected that as in the early years like 2009-2015, the number of publications was minimal in consequence producing no citation. We mentioned previously that the growth of publications started in this area in 2019. From figure 3, it can be observed that the number of citations was also taken a hike from the year 2019 onward, although the year 2018 also received a moderate number of citations. However, the year 2020 was a breakthrough for this particular research topic. A total of 220 citations was found in the year 2020, and the number of citations was approximately triple of the previous year(2019). It was expected that, as the year 2021 has received the highest number of publications in this domain, this year also has the highest number of citations which was found to be 262. This citation analysis shows that this area is getting widely accepted among researchers worldwide, and researchers are very engaged in integrating AI and ML with a circular economy.

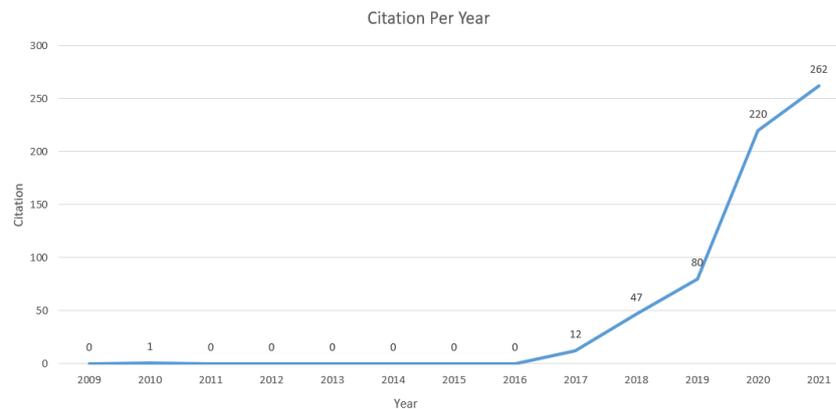

**Figure 3.** Number of Citations Per Year

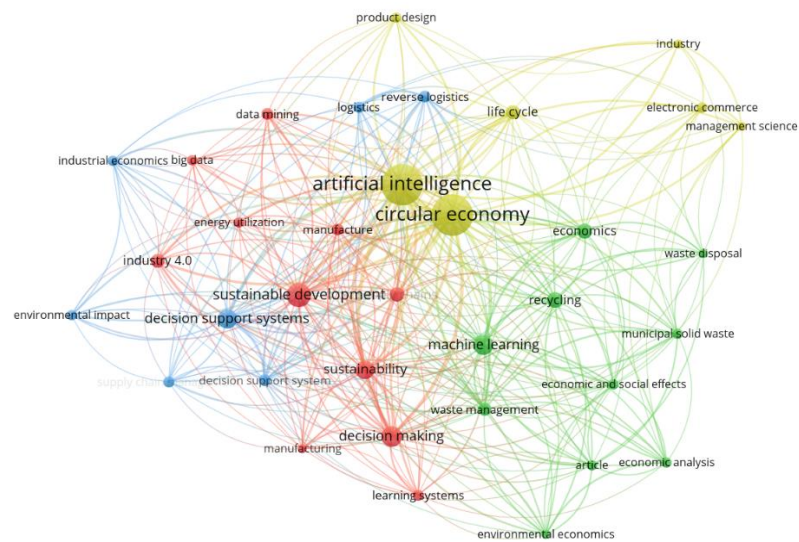

**Figure 4.** Key-Word Co-occurrence Map

### 4.1.4. Keyword Cooccurrence

Keywords are the words mentioned after the abstract section in an article and some other index terms associated with the document. These are the principal terms used throughout the whole article, and these terms are directly involved with the perspective of that article. We used VOS viewer software to have a visual overview of the co-occurrences of keywords among those 104 documents. The total keyword count was 1184. We set the minimum number of occurrences of a keyword to be 5. That means the keywords that occurred five times in all the documents would have appeared in our representation. We found that a total of 35 keywords met our threshold. Figure 4 represents our findings in a co-occurrence graph, which also shows the other key terms involved. From Table 3, we can see the number of occurrences of these

keywords. As expected, artificial intelligence and circular economy are the most occurring keywords, and both are mentioned 68 times. After that, sustainable development, decision-making, and machine learning are in third, fourth, and fifth place.

**Table 3.** Key-word occurrence List (Top 15)

| Keyword | Number of Occurrences |
|---|---|
| Artificial Intelligence Circular Economy | 68 |
| Sustainable Development | 27 |
| Decision Making | 21 |
| Machine Learning | 19 |
| Decision Support Systems | 18 |
| Sustainability | 17 |
| Recycling | 13 |
| Economics | 12 |
| Supply Chains | 11 |
| Industry 4.0 Life Cycle | 10 |
| Waste Management Data Mining Decision Support System Manufacture Reverse Logistics | 8 |
| Learning Systems Logistics Supply Chain Management | 7 |

## 4.2. Country

A total of 39 countries were found to be the origin of the published documents. This section describes country-wise participation in publishing documents on our topic of interest, AI integrated Circular Economic. We highlight the total number of publications of a particular country, total citation per country, country-wise average citation and collaboration among different countries in the following sub-sections.

### 4.2.1. Total Publication

Table 4 illustrates that China is in the first place, having 23 documents published, followed by the UK, Germany, and the USA. The UK is in the second position with 14 documents, while Germany obtained the third position with 11 documents and USA acquired the fourth position with nine documents published. India, Italy, and the Netherlands published seven documents, putting them in the fourth position. Finland, France, and Russia are also not very far from this competition. They all have six documents allocated to the next position on the table.

**Table 4.** Country Total Publication and Citation List (Top 20)

| Country | Documents | Documents Percentage | Citations | Average Citations |
|---|---|---|---|---|
| China | 23 | 15.54 | 28 | 1.22 |
| UK | 14 | 9.46 | 293 | 20.93 |
| Germany | 11 | 7.43 | 310 | 28.18 |
| USA | 9 | 6.08 | 11 | 1.22 |
| India | 7 | 4.73 | 86 | 12.29 |
| Italy | 7 | 4.73 | 34 | 4.86 |
| Netherlands | 7 | 4.73 | 18 | 2.57 |
| Finland | 6 | 4.05 | 258 | 43.00 |
| France | 6 | 4.05 | 26 | 4.33 |
| Russia | 6 | 4.05 | 2 | 0.33 |
| Poland | 5 | 3.38 | 18 | 3.60 |
| Australia | 4 | 2.70 | 11 | 2.75 |
| Greece | 4 | 2.70 | 27 | 6.75 |
| South Africa | 4 | 2.70 | 25 | 6.25 |
| Austria | 3 | 2.03 | 9 | 3.00 |
| Spain | 3 | 2.03 | 8 | 2.67 |
| Canada | 2 | 1.35 | 1 | 0.50 |
| Czech Republic | 2 | 1.35 | 18 | 9.00 |
| Norway | 2 | 1.35 | 13 | 6.50 |
| Romania | 2 | 1.35 | 0 | 0.00 |

#### 4.2.2. Total Citation

Although China is in the first position in the total number of publications, they are very far from the first place when we consider total citations. Here Germany is at the top, getting 310 citations only from 11 documents. The UK takes second place with a competitive score indistinguishable from Finland. The UK got a total citation of 293, whereas Finland is 35 citations shorter than Germany. The rest of the countries are very far from this trio.

#### 4.2.3. Average Citation

Finland showed an extraordinary performance to acquire the first position here. The average number of citations per document is 43.00 for Finland. Germany and the UK secured the second and third positions with an average citation of 28.18 and 20.93 per document. On the other hand, China did an abysmal job here. Their average citation is 1.22 only. Having many documents published but scoring poorly with citations certainly raised concerns for a particular country. China, in this case, has a concerning average cite-score.

#### 4.2.4. Country Collaboration

From 39 countries contributing to the research on AI-aided circular economy, 32 countries participated in international collaboration, measuring 82% of the total countries. Collaboration between countries and researchers is a commendable sign in this field as a country itself cannot implement circular economy or sustainability alone as it is a global phenomenon. Countries worldwide need to come ahead to implement a complete circulation of the economy, and for this reason, the international collaboration of the researchers is crucial. Figure 5 depicts the countries' collaboration where cluster centroids are countries publishing more documents while also leading in collaboration. The cluster elements are the countries that intensely collaborated with the country in the centroid position. The size of the circle represents the total number of documents published from a country, and the thickness of the linking line between countries represents how stoutly they collaborated. As we can notice from figure 5, the set of 32 countries has produced 8 clusters. These clusters are explicitly indicated in the figure by various color codes. Sequentially, cluster 1-8 is led by India, Germany, South Africa, Netherlands, United Kingdom, Italy, China, and Russian Federation. From the size of the circles, it can be identified that the United Kingdom, China, France, and Australia are the most collaborating countries where the United Kingdom collaborated with 13 countries and China, France, Germany, and Australia all collaborated with eight countries. India collaborated with six countries but worked as a linkage between four major collaborative countries: the United Kingdom, China, France, and Australia. Italy and the Netherlands also played their role by collaborating with five countries. In contrast, Russian Federation does not seem to be interested in a collaborative approach.

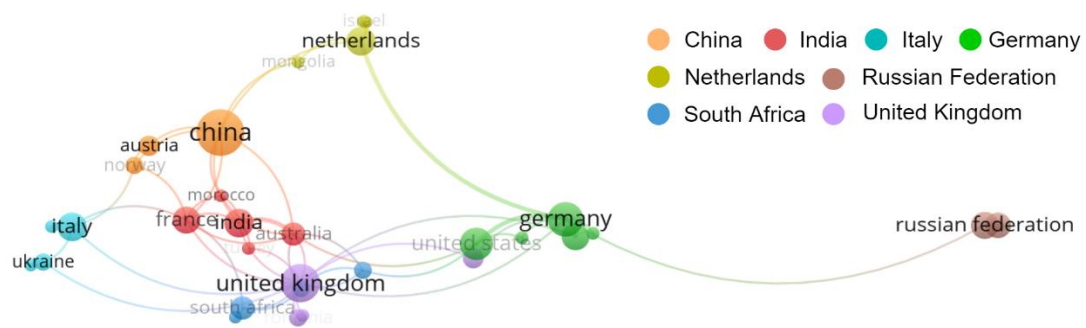

**Figure 5.** Country Collaboration Map

### 4.3. Subject Area

Figure 6 represents the subject area related to this study. Our main keywords (AI and ML aided circular economy) are often related to computer science, engineering and environmental. The subject area engineering placed top in the list, although both engineering and computer science marked almost the same involvement in this study. Ecological science is involved in 34 studies recorded. It is evident that the goal of circular economy is to aid the betterment of the environment, environmental science, and energy will appear among the top. However, we discovered that many areas like Business Management, Social Science, Decision Science, and Economics are significantly involved in this field of research. We also observed some rare fields like Psychology, Physics, Astronomy, Neuroscience, and Biochemistry, which makes the area of our study more versatile. The figure below (figure - 6) shows that many emerging domains are blended with the circular economy nowadays. Researchers globally show some creativity and expose some novel ideas to integrate new areas with the circular economy. Soon, these unique ideas and unification of the fields will result in a real-life solution to efficiently implement the circular economy in a more ubiquitous synopsis.

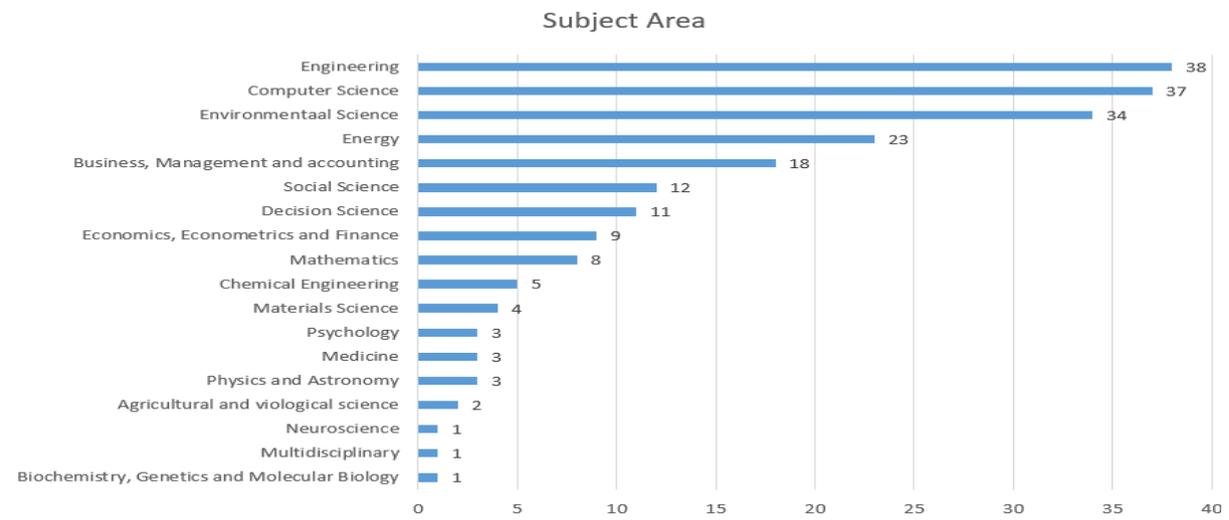

**Figure 6.** Subject Area of this Study

## 4.4. Author

### 4.4.1. Active Authors

This section describes the highly active authors who contributed the most in this domain. First, the authors having at least two published documents have been considered due to fewer documents found. However, as this area is considerably new, most authors have only one publication, and the authors who have more than one publication have not been cited in most cases. So, we decided to add those authors who have a single publication but more than ten citations. We found a total of 56 authors that matches our criteria to find the active authors. Table 5 delineates these authors with their total number of documents and citations. Among the 56 authors, only 14 authors have more than one publication. The rest of the authors have a single publication that received substantial citations. Moreover, we observed that Ramakrishna S. has the greatest number of publications. He has three publications with eight citations, with an average of 2.67 per publication. It is the highest among the other authors. The rest of the 13 authors have two publications, and among them, Zijm H. & Klumpp M. are in the top with 15 citations, followed by Ghoreishi M. & Happonen A., who have ten citations in two documents. Matta N., Reyes T., & Bratec F. have two publications but have not been cited yet.

The rest of the authors with two publications have at least more than one citation. Then comes the authors who have one publication but acquired the most significant number of citations. D'Amato *et al.* [57] showed an extraordinary performance here. These nine authors worked for a single publication, and they got 244 citations from that publication, which is the most impactful work in this area. Rajput S. & Singh S.P. also has a single publication and got 83 citations, the second most impactful work. Another impactful publication was done by Reuter M.A., which received 50 citations. The rest of the authors listed in the table (Table 5) have at least ten to 18 citations in a single document.

**Table 5.** Author's document and Total Citations in this Domain

| Authors Name | Document | Citations | Average Citations |
|---|---|---|---|
| Ramakrishna S. [9,43,44] | 3 | 8 | 2.67 |
| Zijm H., Klumpp M. [45,46] | 2 | 15 | 7.50 |
| Ghoreishi M., Happonen A. [35,47] | 2 | 10 | 5.00 |
| Wu Y.[48,49] | 2 | 5 | 2.50 |

| Chen W.[42,50] | 2 | 3 | 1.50 |
| Makarova I., Shubenkova K., Pashkevich A. [51,52] | 2 | 2 | 1.00 |
| Zhang X. [53,54] | 2 | 1 | 0.50 |
| Matta N., Reyes T., Bratec F. [55,56] | 2 | 0 | 0.00 |
| D'Amato D., Droste N., Allen B., Kettunen M., Lähtinen K., Korhonen J., Leskinen P., Matthies B.D., Toppinen A. [57] | 1 | 244 | 244.00 |
| Rajput S., Singh S.P. [33] | 1 | 83 | 83.00 |
| Reuter M.A. [58] | 1 | 50 | 50.00 |
| Vondra M., Touš M., Teng S.Y. [59] | 1 | 18 | 18.00 |
| Bag S., Pretorius J.H.C., Gupta S., Dwivedi Y.K,Bianchini A., Rossi J., Pellegrini M. [60] | 1 | 16 | 16.00 |
| Charnley F., Tiwari D., Hutabarat W., Moreno M., Okorie O., Tiwari A.,Kokkinos K., Karayannis V., Moustakas K. [61] | 1 | 14 | 14.00 |
| Borowski P.F. [62] | 1 | 11 | 11.00 |
| Vlachokostas C., Achillas C., Agnantiaris I., Michailidou A.V., Pallas C., Feleki E., Moussiopoulos N.,Magazzino C., Mele M., Schneider N., Sarkodie S.A. [63] | 1 | 10 | 10.00 |

### 4.4.2. Author's Dominance Ranking

The author's dominance factor (DF) is an assessment of an author's contribution to a publication. It is a proportion that indicates how often in a multi-authored articles have a particular author has been listed as the first author. DF factor have been popularly used by many bibliometric studies as a metric to measure contribution by an author [64-66]. It is a score  between 0 to 1 where the number of first-authored papers of a particular author is divided by the total number of multi-authored papers by that author. The articles in which an author is the sole author, have not been included in this measure, since, the ratio will always yield "one."  The ADF (Author Dominance factor is measured by the following equation:

**ADF = (Total first-authored papers / Total multi-authored papers)**

Table 6 depicts the dominance ranking of the authors. Here we considered only the authors who have at least two publications. Zhang X. has been included in this list because the author has two publications, but one is a single-authored paper. We had to omit that paper from the list as per rules, but the author is there for another multi-authored paper. As we can see from the table, five authors from the list scored the maximum DF factor, which is "1" because of the first authorship on all of their papers. Nevertheless, Chen W. and Ramakrishna S.  scored 0.50 and 0.33, respectively, as they have both single-authored and multi-authored papers. Wu Y. scored zero on this list because he is not the first author in his publications.

**Table 6.** Author Dominance Ranking

| Authors Name | Rank By DF | DF | Multi Authored | First Authored |
|---|---|---|---|---|
| Ramakrishna S. | 7 | 0.33 | 3 | 1 |
| Zijm H. | 4 | 1.00 | 2 | 2 |
| Ghoreishi M. | 1 | 1.00 | 2 | 2 |
| Matta N. | 2 | 1.00 | 2 | 2 |
| Makarova I. | 3 | 1.00 | 2 | 2 |
| Wu Y. | 8 | 0.00 | 2 | 0 |
| Chen W. | 6 | 0.50 | 2 | 1 |
| Zhang X. | 5 | 1.00 | 1 | 1 |

### 4.4.3. Author's Keywords

This section presents the technical keywords in this area which the authors have used only after the abstract section. Table 7 depicts the keywords frequencies. The most occurring keyword is circular economy, followed by artificial intelligence and machine learning.

**Table 7.** Authors Keywords in this Domain

| Keywords | Occurrences |
|---|---|
| Circular Economy | 50 |
| Artificial Intelligence | 22 |
| Machine Learning | 14 |
| Sustainability | 11 |
| Industry 4.0 | 9 |
| Decision Support System Reverse Logistics | 7 |
| Recycling | 5 |

| Sustainable Development | |
|---|---|
| Big Data | 4 |
| Artificial Neural Network | |
| Circular Design | |
| Decision Support Systems | |
| Deep Learning | |
| Digital Economy | 3 |
| Logistics | |
| Municipal Solid Waste | |
| Remanufacturing | |
| Waste Management | |
| Bioeconomy | 2 |

Sustainability comes after that, and it occurs 11 times with sustainable development, which occurs five times. Both are roughly the same concept and reflect the same idea. One of the significant concepts of the circular economy is to bring sustainability or sustainable development where it is tried heart and soul to do as little harm as possible to nature while satisfying human advancement aims. Decreasing all types of pollution and recycling the used items is also vital for aiding these ideas. That is why recycling and reverse logistics are also important keywords used by the authors. Furthermore, the fourth industrial revolution, decision support system, and circular design also came on the list as they are closely associated with sustainable development and a circular economy. Figure 7 depicts the co-occurrence of the author's keyword. From this figure, we can visualize the linkage between the keywords and how each keyword is associated with the other.

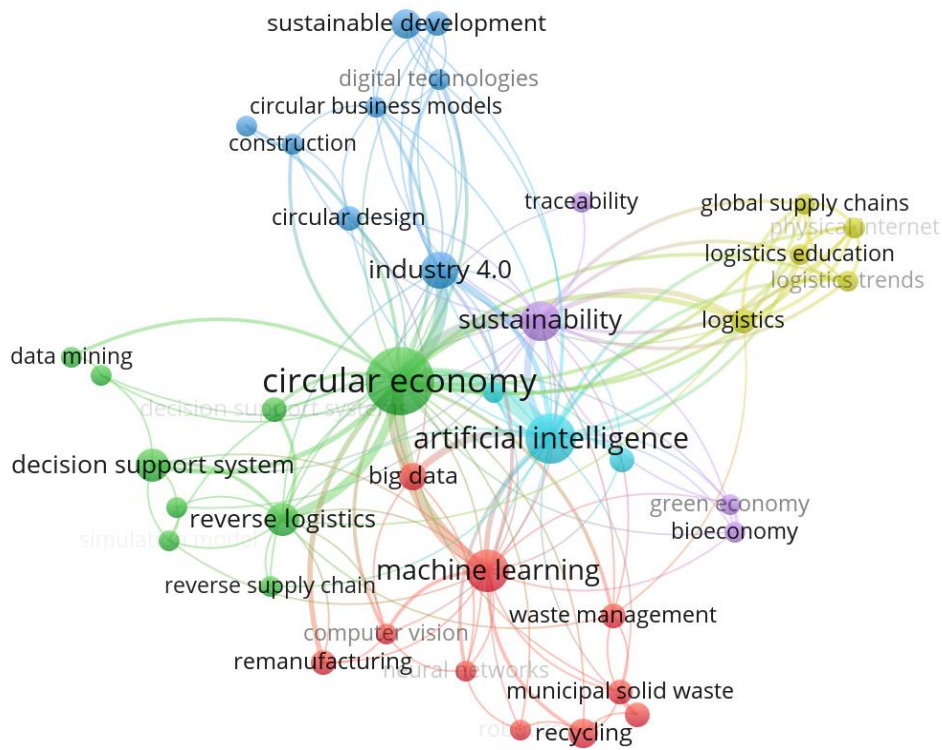

**Figure 7.** Co-occurrence Network of the Authors' Keywords

## 5. Systematic Literature Review

For the systematic literature review(SLR) section, we have manually divided the documents into six categories according to their subject area to find out the current research trends in this domain. The names of the categories are sustainable development, reverse logistics, waste management, supply chain management, recycle & reuse and manufacturing development. Some of the documents have fulfilled the criteria to stay in multiple categories. Each category has been further discussed with the documents related to that specific category below.

**5.1. Sustainable Development**

Sustainable development(SD) is one of the central pillars of the CE system. SD is a collection of policies aimed at capturing people's progress while preserving the ability of nature to provide raw materials and ecosystem services [67]. To bring circularity in the economy, the ideas related to CE need to satisfy sustainability goals. In our investigation, we discovered a significant number of articles in which researchers use SD with the assistance of AI to bring about circularity in the economy. Researchers have explored the area of heat recovery [59], power optimization [27], carbon footprint reduction [28], e-waste management [68], sustainable product design [47], and eventually building sustainable cities [69] to ensure sustainable development leading to CE. Details on these works have been discussed in the subsequent paragraphs.

Marek *et al.* [59] demonstrated an evaporator system implementation to propose a sustainable liquid digester treating mechanism based on the central idea of CE. Their research employs a novel architecture of tree-based machine learning algorithms namely decision tree and random forest to investigate the techno-economic viability of adopting vacuum evaporation systems to encapsulate waste heat from connected heat and power units in biogas facilities. The heat produced during production is completely wasted in linear economy and encapsulating this heat can aid some other production bringing circularity. They also employed the Monte-Carlo simulation method to produce multiple evaporation system options to make the model more robust. The best system was picked using artificial neural networks and decision tree classifiers [59]. Their findings revealed that integrating the evaporation system is beneficial when transportation costs and incentives for coupled heat and power units are high but specific investment prices are low. Wang and Zhang [27] introduced an analytical energy control technique based on several costing frameworks for power and cost optimization using a generalized ML model termed modular neural network. This model can estimate the optimum idleness of an enterprise using power optimization and cost factors and perform techniques to optimize profitability that can also support sustainability in supply chain management.

The overall volume of greenhouse gases (primarily carbon dioxide and methane) generated by a structure during its life cycle is acknowledged as its total carbon footprint. Reducing carbon footprint is one of the most critical issues in CE as it instantly affects environmental sustainability. . In regenerative architectural design of building construction, Mazurek *et al.* [28] employed machine learning methods, convolutional networks, and parametric design to minimize the carbon footprint along the construction. A regenerative architecture can take the advantage of carbon footprint optimization to bring CE by reusing the structural materials circularly. [28] Their three research steps culminated in the creation of a functioning machine learning model that might help architects in the early stages of constructing environmentally conscious design models. Based on limited data, the final research created a trained Neural Network capable of estimating the complete Carbon Footprint of a design proposal. 53.6 million tons of e-waste are produced each year globally, where only 9.3 million tons of waste are processed. At the same time, the rest heavily contribute to e-waste generation [70]. Several efforts have been seen [68,71] to process this waste, and waste generation has also been reduced.

As the nations are adopting the SD policies, it is essential to know the progress of the nations in this race. Chang and Lee's work applied one of the unsupervised ML techniques known as the k-means method to discover a relationship between five identified capital resources from a dataset gathered from 32 OECD (Organization for Economic Cooperation and Development) countries. They built an SD progress index using forest-based classifications and regression using the connections [72]. Their study provides countries with a clear perspective on regulating resource demand and economic development. On the other hand, Li *et al.* [42] proposed a multi-level knowledge-based architecture that can dynamically monitor the sustainable development of the regions. This entire system was created by combining artificial intelligence, comprehensive management science, and economic mathematics to dynamically observe and judge the whole operation, which also works as a decision support system for the management. Other than this, some researchers focused on the topics which fall into different categories but are indirectly responsible for supporting sustainable development, for instance, sustainable household waste management [73], sustainable supplier selection [7], integrated manufacture through sustainable technology[48], sustainable e-waste collection[68], sustainable wastewater treatment [74].

### 5.2. Reverse Logistics

Reverse logistics is the process of returning faulty products or any part of the product (that is not further usable) to the manufacturer using a reverse supply chain system for re-manufacturing (through refurbishment) or recycling (if unusable). This process is directly related to recycling. The reverse logistic system has gained popularity due to the severe extension of the e-commerce network [75]. In the circular economy concept, reverse logistics is probably the shortest and the most cost-effective way to complete a product's life cycle as a faulty product gets back to the producer within a short time completing its life cycle. Efforts towards applying the reverse logistics practice with the help of artificial intelligence to establish circularity in the economic system have been seen [29,76,77]. Several studies on reverse logistics to achieve CE has been found where researchers focused on faulty products returning tactics [52], optimization of e-waste (concentrating on a mobile phone) recycling [78], assuring product quality to better understand its reusability [79] and the overall automation of reverse logistics process [29].

Wilson *et al.* [76] presented a study where they tried to explore the scope of AI for the reverse logistics system. A reverse progression of a product indicates remanufacturing, reusing, refurbishing or recycling. Their study found that AI renders tremendous additional privileges with all the tasks related to the reverse logistics process. Still, the different functions of reverse logistics may require diverse forms of AI. A study done by Schulter *et al.* [29] tried to apply machine learning and image recognition system to automate the identification, inspection, sorting, and reprocessing of the products' re-manufacturing after each life cycle of a product. Their study showed that intelligent AI algorithms and constant digitalization could analyze complicated operations like reverse logistics or recycling management more straightforward for the employees. Zheng *et al.* [77] designed a comprehensive evaluation index system for the reverse logistics-related operations of municipal solid wastes with logistics flow. Solid wastes management can contribute to the circularity in a product's life cycle. This system finds the bottleneck of the whole system. Using that bottleneck can improve the effectiveness of the reverse logistics operations for optimizing and enhancing the management process and finally improving the overall performance of the city's solid waste management.

The automobile industry is one of the largest sectors in the global market. In transforming to a circular economy, the reverse delivery process of automobile faulty spare parts is an important issue. Makarova *et al.* [52] developed a decision support system to predict, prepare and arrange the delivery of new spare parts and return faulty parts on the same trip for re-manufacturing. Re-manufacturing of a product prolongs the life cycle of the spare parts and hence contributes to CE. An architecture was developed by them which can efficiently minimize the empty runs of the delivery vehicle and work as a reverse logistics method in the circular economy context. The study of Song *et al.* [78] showed an optimized reverse supply chain network for waste mobile phone recycling as it brought immense harm to the environment and resources. These resources can be recycled, reused and remanufactured to increase the life cycle. Their model is based on fuzzy chance-constrained programming, transformed into equivalent fuzzy variable forms. Their study also proposed an algorithm called particle swarm optimization with a dynamic changing inertia tool which results in an optimal layout of transportation path and facility nodes for a province of China. Mobile phone supplier enterprises can adopt this reverse logistics system to reduce the waste of mobile phones, which impairs the environment. To acquire a complete circular economy, the reuse of products is vital. Reverse logistics systems and closed-loop supply chains are considered critical elements for this economic transition. The study of Lechner and Reimann presents a nonlinear optimization model with interconnected techniques based on mathematical modeling for acquiring used products, assorting for product quality assurance, and reprocessing disposal [79]. From the perspective of an independent reprocessing business, the system can function as an integrated decision support system in reverse logistics and closed-loop supply chains.

### 5.3. Waste Management

Waste management is the process or activities related to handling the waste to make it less harmful to the environment. The waste management process consists of different activities like collection, transportation, monitoring, processing, and waste disposal. In the context of CE, better resource management and reducing the amount of more waste to be produced is one of the main factors. Our study found several articles where researchers worked on improving the management of different types of waste, including solid waste [80,24], bio waste[633], liquid waste[74,81] and e-waste[68] with the help of different AI techniques. Articles focusing only on municipal waste management were also found [24,30,82,83]. For instance, Rosecky' *et al.* [30] investigated several factors that impact municipal waste generation at different territories like the regional, micro-regional or municipal area in the Czech Republic. They used mixed forms of linear and tree-based ML models like linear regression, regression trees, and random forest and found that different combination of ML models serves divergent purposes. To illustrate, the linear regression model allows a trade-off within accuracy and interpretability. Random forest is good at prediction with high accuracy. Accurately forecasting waste generation can help managing waste generation at the first place, reducing the amount of waste and finally to recycle. Fasano *et al.* [82] did almost identical types of work. They built deep learning-based architectures to examine the variable that most affects the production of waste, separate collection, and management costs of municipal solid waste. They used 102 variables, and different combinations of variables were used for predicting individual tasks. Different combination of layer setup was also a subject of their study [82]. Their study experimentally proves that these models could help efficient municipal solid waste management services from diverse perspectives.

Apart from labor-based waste management, the use of robotics has automated the process of waste management, saving valuable time and resources. Wilts *et al.* [24] suggested an AI-based robotics sorting system of bulky municipal solid waste and a full-scale waste treatment plant. Alonso *et al.* [23] used convolutional neural networks and image recognition to automatically separate contrasting waste materials such as glass, plastic, paper, and organic elements for further recycling. Another study was done by Paul and Bussemaker [83]. They implemented a web application tool using web-GIS that renders essential information about the supply chain of an area to evaluate the waste valorization suitability of a distinct location [83]. This web application can serve as a decision support system for planners and municipal solid waste authorities to mitigate future problems in handling the growing volume of waste. Ramchandani *et al.* [18] presented a case study to manage the packaging waste of large multinational enterprises by integrating blockchain and AI. Their study found that large global enterprises are not concerned about their packaging waste; however, the problem can be solved by using the right technology and providing the appropriate motivations to the involved actors. Papagiannis *et al.* [73] showed the household waste management approach using k-means clustering for estimating urban residual waste based on its generation level.

The concrete demolition waste management approach was shown by Sobotka and Sagan, where they used a mathematical simulation model as a backbone [80]. Their approach works as a decision-making tool while selecting technological and organizational resolution in concrete waste management. Vlachokostas *et al.* [63] proposed a generic model using a decision support system that can concurrently examine environmental, economic, and social criteria to generate three different units of alternative biowaste treatment (small, medium, or large-scale) to enhance the production of bioenergy and bioproducts. Bio energy is clean and renewable which completely satisfies the idea of CE. Their experiments showed notable results, verified in the region called Serres situated in Greece. According to their investigation, the model proposed by them was generic enough to be adopted worldwide to solve related issues in supply chain management. With a slight modification of the threshold values, their model can aid in locating the optimal location of warehousing, collection points, and sorting centers.

Wastewater treatment is another critical issue in waste management, and two articles extensively focused on this topic [74,84]. The study of Sundui *et al.* [84] exhibits an extensive review of machine learning and deep learning algorithms on the biological wastewater treatment process to recover nutrient and biomass production from municipal wastewater. Their study found that machine learning algorithms and artificial neural networks are more popular considering deep neural networks(CNN, RNN). Finally, another gloom-ridden type of waste known as electronic waste is covered by only one article. Nowakowski *et al.* [68] developed an online-based sup-porting system to collect electronic waste, where they used the Harmony Search algorithm to optimize the route of the waste collection vehicles. Their study also presents some selected algorithms' performance better than other AI algorithms. Their presentation also included a novel design of vehicle bodies to improve the efficiency of waste loading and packing, especially in densely populated areas.

#### 5.4. Supply Chain Management

Supply chain management manages the flow of products and services from the point of origin to the point of consumption, including the transportation and storage of raw materials, working process inventories, and completed commodities. In the circular economy context, supply chain management is regarded as a critical issue to focus on. A few articles contributed directly or indirectly to improving supply chain management so that it promotes CE. Kazancoglu *et al.* [85] tried to find the barriers related to the dairy supply chain that restrict achieving circularity in that field. Their framework is based on six barriers and twenty-seven sub barriers where they pro- posed different types of big data solutions using two fuzzy-based methods. Their results show that economic obstacle is the most weighted barrier. Machine learning & data mining are the most efficient big data solutions to achieve circularity in dairy supply chain management. An IoT-enabled decision support system-based innovative Circular economic business model was shown by Mboli *et al.* [86]. The system can track, monitor, and analyze the product in real-time by focusing on the residual value [86]. Adopting this model will be a great help for the suppliers to maintain a circular supply chain in the market.

In CE, the reverse supply chain process plays an important role by reducing the amount of new waste generation and recycling the existing products. We found two articles that discussed the reverse supply chain [78,87], where authors proposed a design for a reverse supply chain network for waste mobile phones[78] and for automobile parts recycling [87]. Makarova *et al.* [52] presented a decision support system for the supplier of automotive spare parts. They reversed the supply of faulty parts back to the manufacturer in the same run, ultimately reducing empty runs of the supplier vehicles. A study done by Vlachokostas *et al.* [63] suggests using a decision support system to implement the alternative biowaste treatment. The study done by Alavi *et al.* [7] came up with a different approach where users can find suitable and sustainable suppliers based on some customized social, economic, and circular criteria. Their approach uses a fuzzy logic-based method to find the most suitable and sustainable supplier to obtain supply chain profitability [7].

#### 5.5. Recycling and Reuse

The concept of recycling and reuse is often bewildering. Recycling is the process of collecting waste materials and breaking them down into building blocks for again turning them into products, whereas reusing is the process of taking old items and using them. Both are excellent practices to reduce waste and keep the environment cleaner in the circular economy concept. The advancement of technology has enabled us to adopt advanced systems such as robotic recycling to achieve CE. The proposed design of Alonso *et al.* [23] can separate waste materials such as plastic, paper, glass, and organic materials for recycling and reusing purposes using convolutional neural networks and image recognition systems.

The assembly and disassembly-related features of a product impact its recycling and reusability of it. The easier a product is to disassemble, the easier it is to recycle and reuse. A robot-based disassembly system is shown by Poschmann *et al.* [88]. They used TensorFlow object detection API using faster RCNN and Inception ResNet V2 structure. The whole system can use the life cycle data of products to decide the actual degree of disassembly for the best end-of-life utilization of that product [88]. . Stavropoulos *et al.* [89] presented a study using a modular design support system that works with a particularized design point for any multi-material element in automated production lines and uses a decision support system to provide the users with assembly or disassembly-related factors and circumstances. Their research focuses on product reusability and recycling by bridging the gap between multi-material component designers and past knowledge of disassembly characteristics.

Spare parts of everyday used machines are a massive contributor to spare-part waste. Once a spare-part is damaged, it may not be used in another machine and becomes a waste. Fan and Cai [87] designed the automobile used parts recycling cost prediction model for suppliers using ARMA analysis. This model will also work as a data mining tool for the enterprises to take the decision of recycling of a specific automobile used part. Using supervised machine learning approaches, Rakhshan *et al.* [8] developed a probabilistic predictive model to examine the economic reusability of the building's load-bearing components using supervised machine learning methods. They applied powerful machine learning algorithms to determine the economic aspects impacting the reuse of structural parts of buildings in the first phase of their system. They devised a set of principles for practitioners to successfully quantify the economic reusability of structural components in the second segment [8].

Product recycling is undoubtedly commendable from an environmental perspective. Still, if we consider it from an economic perspective, it is possible to cause unnecessary financial loss. Dent *et al.* [90] propose an experimental model to detect the economic utility of product recycling using machine learning algorithms. They use a deep neural network and support vector machine algorithm to distinguish between profitable and non-profitable end-of-life recycling products. Their proposed method can also be adopted in other circular economy regions for a more reliable decision-making process [90]. The work of Huang [91] proposes a unique perspective to the concept of reusing. He initiated a new designing tool that uses a machine-learning-based searching system to design a completely new structure from existing waste plastic materials using reinforcement learning and machine vision. This tool can help classify and convert random pieces of wasted faulty plastic into a new form. Adopting this system more broadly may help reduce plastic-based waste by recycling the trash materials into a new helpful form.

### 5.6. Manufacturing Development

Manufacturing development, also known as manufacturing process development, improves product design, digitalizes the conventional approach, integrates or brings advanced technologies in the product manufacturing phase. Bringing improvement to the product manufacturing stage may help in the product's supply chain, recycling, or reverse logistics process, which can directly or indirectly promote the circular economy [92]. Although this is an essential concept in the circular economy, a handful of articles focused on this topic were found. Su *et al.* [48] presented an integrated manufacturing system that incorporates the whole product development method from raw materials to end-of-life treatment. Three circular economy-based business concepts, such as the co-creation of products, sustainable consumption, and collaborative recycling guidelines were also investigated in their study. These models are also supported by various techniques such as information technology, traceability, eco-point approach, and online data-mining tool to know the details of a specific product and make it less harmful to the environment.

Using machine learning, a unique technique for developing and accessing a novel waste heat recovery system for carbon fiber production has been reported by Khayyam *et al.* [9]. To forecast energy usage, their conceptual design approach uses an artificial neural network and nonlinear regression. Utilizing this prediction can effectively reduce the cost of carbon fiber production and carbon footprint from the whole process. Ghoreshi and Happonen [47] performed an extensive literature review of the design of circular material and integrations of AI and CE with smart production. Based on this theoretical study, they also proposed a framework for the uses of AI in the design of circular products. Following their framework, manufacturing enterprises can integrate AI techniques into CE in the product design phase that may help develop product sustainability cost-effectively.

Chen *et al.* [93] addressed the environmental cost control mode and its challenges, as well as the link between environmental cost control and the value chain. To promote a circular economy, it is crucial to do as little harm as possible to the environment. As a result, the control of business environmental protection costs is the primary concern for modern enterprises. This work efficiently designs manufacturing enterprises' environmental cost control system by applying a decision tree algorithm [93]. The enterprises can effectively manage the cost of the environment by utilizing this model compared to the conventional business model. From the context of circular economy, Schluter *et al.* [29] showed that after completing the life cycle, old items might be returned to a manufacturer for various processes like classification, inspection, sorting, and reprocessing for remanufacturing purposes. Using advanced AI technology, remanufacturers may automate these processes. They also developed an architecture using applied AI techniques to convert these complex tasks into simple decisions to complete them efficiently[29].

### 6. Discussion and Research Gap

This study utilizes a bibliometric analysis followed by a systematic literature review to investigate AI-integrated systems' progress and current research trends to achieve CE. In this section, we have presented a detailed analysis of current research included in our study to explore and find out the specific research topics that need further exploration. For the bibliometric analysis, the data has been organized into four sections: country, author, document, and subject area, with multiple subsections to find out the growing trends, country-wise research contribution, and prolific authors in the subject area and collaborative research among different countries. Total 32 countries participated in international collaboration. India has collaborated the most in research on the field of AI-aided Circular economy. China has the greatest number of publications with minor average citations. Germany is ranked first in the total number of citations. Finland has a higher average citation than others. Most of the papers of our subjective topic, are from the environmental science, computer science, and engineering sec- tors as suggested by the name of the field itself (AI-aided CE). Seeram Ramakrishna is the highest contributing researcher as determined by our study. 2019 was the year when the research efforts on this field gained momentum, and this field of study took a hike in the following years.

The circular economy is a broad topic, and for the systematic literature review, we divided the documents into six different categories to explore the field of AI-aided CE from different perspectives. The categories are sustainable development, reverse logistics, waste management, supply chain management, recycling & reuse, and manufacturing development. This study discovered that researchers tried to blend sustainability from the application level [27,28,59] to the assessment level [42,72,94] to promote a circular economy. However, sustainability is not solely enough to bring about CE as it is a collective approach requiring several factors to work together. Kadar *et al.* [95] discussed that sustainability could not be a long-term solution, making things less harmful to the environment but not completely regenerative. An efficient regenerative economy needs a lot of data and new techniques of AI in all steps of the ecological system to achieve CE. Cetin *et al.* [17] explored the most appropriate digital technologies to enable circular economy in the built environment and found that AI is particularly suitable for advanced data-driven regenerative design, where the built environment refers to the human settling and its surrounding.

AI can also help to promote the reverse logistics infrastructure for a closed-loop environment by developing the processes to classify faulty products, remanufacture components and recycle materials3. For instance, the study of Wilson *et al.* [76] presented that different form of AI renders tremendous additional privileges with different functions related to reverse logistics. In this particular instance, researchers have to be aware of finding the best suitable AI algorithms for the diverse purpose of reverse logistics systems related to the circular economy. Another crucial factor in CE is waste management for accomplishing better resource management and more waste prevention. Most of the articles in this category focus on municipal solid waste management [24,30,82,83] and liquid waste [74,81]. However, according to Ellen Macarthur Foundation, the potential value of AI in improving the design of food and electronics waste is jointly equivalent to almost 217 billion a year in 20303. Our study found only two articles regarding this issue, where one article focused on bio-waste [63], and another was focused on e-waste [68]. This gap has to be filled, and researchers need to come forward to find out more generative designs in these two waste management sectors to utilize the best use of AI to obtain this tremendous potential benefit.

Waste management and prevention require a reliable supply chain, especially a food supply chain network. In our study, we discovered less focus from the researchers on this issue. Although the current global food production system is growing more than enough food to feed the entire population, almost one-third of the food is lost during consumption and due to inefficient supply chain management in the food sector3. In the context of a circular economy, food wastage must be checked, and for this purpose, AI can be applied to forecast the supply and demand of a particular area using previous and real-time data to improve the efficiency of the supply chain. A casual network has been developed by Gue *et al.* [19] using a fuzzy cognitive map for urban circular economics using six drivers and one goal to understand direct and indirect influencers of urban circularity. Researchers can focus on building the same type of global supply chain network map to find the barriers related to the supply chain and restructure the supply chain accordingly to find the optimized route, which may also promote better utilization of food storage and freezing facilities.

---

[3] Ellen MacArthur Foundation, Artificial intelligence and the circular economy - AI as a tool to accelerate the transition (2019). http://www.ellenmacarthurfoundation.org/publications

Technologies like blockchain and IoT integrated with AI can help traceability and transparency in the supply chain [96,97]. According to Cetin *et al.* [17], blockchain is an advanced technology to manage complex information networks in the supply chain, control the cash flow through a smart contract and act as a shared database between stakeholders. Although IoT, blockchain and smart contract [98] has been applied in real-time tracking, monitoring, and analyzing the product, our study did not find articles that utilize blockchain or smart contracts to manage the supply chain [86]. However, a combination of blockchain and AI has been used to manage the packaging waste of large multinational enterprises by Ramchandani *et al.* [18]. Managing waste through recycling is another critical issue where researchers should be more focused on achieving a comprehensive circular economy. Our study found that researchers were more versatile in this category and focused on integrating AI in the recycling and reusing process from diverse perspectives [23,89,90]. Automatic classification and disassembly processes have been integrated to automate and speed up the process of recycling with the help of image recognition techniques [23,88].

Researchers have also focused on the recycling of automobile used parts [87], reusability of building's load-bearing components [8], the economic utility of product recycling [90], waste plastic materials recycling [91]. However, we did not find any article discussing electronic waste recycling as it is one of the most crucial sectors to focus on. E-waste has a vast market across the globe, and almost 80% of the e-waste is not handled correctly, according to research3. This colossal sector cannot be avoided as it can result in an ultimate environmental infection. Furthermore, our last category is on manufacturing process development. A very minimal number of articles focused on this topic. Researchers focused on different perspectives like integrated manufacturing systems [48], heat recovery systems for carbon fiber manufacturing [9], integrating AI in the design of circular products [47], environmental cost control systems for manufacturing enterprises [93], automating various remanufacturing processes [29]. Many improvements can be added to this sector as it simultaneously promotes reverse logistics, recycling, and supply chain management, which are undoubtedly crucial factors for the transition towards CE. To conclude our findings and research gaps, the major research gaps have been listed below.

- Emerging technologies like blockchain and smart contracts have not been explored.
- E-waste management has not gathered enough focus and needs additional focus.
- We observed limited research on manufacturing process development.
- Alarming lack of research in food supply chain resulting in massive food loss.
- A lot of research focuses on sustainability which cannot be a long-term solution.

## 7. Implications

This study will practically help the stakeholders and future researchers by providing a compiled collection of information from various perspectives. The different research categories presented in our study will help the stakeholders to understand the current trends and shift in technology from a variety of perspectives. The major contribution towards achieving CE from different subject areas, countries and individual researchers have also been delineated. Researchers have shown that, adopting AI techniques assists industry managers to ensure the effective implementation of CE. The fundamental insights gained from this research will also encourage the industry leaders to adopt AI methods for converting their current linear production systems to CE.

Additionally, the study's results will assist industry practitioners in reflecting on different AI methods in the context of CE. Industry practitioners and managers may explore the possibilities of artificial intelligence methods for developing strategies and frameworks for sustainable growth. Product designers may use AI to create more responsive designs, which can aid in waste reduction and management. The supplier can efficiently improve the supply chain with the help of AI for faster supply and reverse logistics support. The manufacturer can also integrate advanced technologies like AI to be more productive in the product manufacturing phase. Some stakeholders, like industrialists, may use AI to bolster circular business models by combining previous-recorded and real-time data from other stakeholders like producers, manufacturers, suppliers, and consumers for process optimization and automated decision-making. Following the research gap, researchers worldwide may also focus on identifying the future opportunities of applied AI within this field. To better illustrate the implications, the major implications have been listed below.

- Knowledge discovered from a wide range of studies has been compiled in this research so that professionals and educators can know the recent trends, research gaps and implications.
- Assist industry practitioners with knowledge on the efficacy of different AI methods in the context of CE and how they can achieve long-term benefits.
- Product designers may use the knowledge depicted by our study to develop more responsive designs, which can aid in waste reduction and management.
- Industrialists may use data-driver AI algorithms to bolster circular business models by combining previous-recorded and real-time data from other stakeholders like producers, manufacturers, suppliers, and consumers for process optimization and automated decision-making.
- The supplier can efficiently improve the supply chain with the help of knowledge discovered by our study to improve supply chain and reverse logistics.

## 8. Conclusion

Circular economy aims to bring out the highest value from raw materials by using them repeatedly. Artificial intelligence is one of the most dominant and potential tools to lead the transition of CE from linear to circular. Our study converges on the research based on artificial intelligence and machine learning techniques in circular economy, incorporating several perspectives: sustainable development, reverse logistics, waste management, supply chain management, recycling & reuse, manufacturing process development. Bibliometric analysis followed by a systematic literature review helps understanding the research trends within this domain by presenting an ensemble of related knowledge. The bibliometric analysis shows that, Germany and United Kingdom have shown keen interest in CE in terms of publication, citation and collaborative research. Unexpectedly, researchers from fields like psychology, astronomy and neuroscience have also contributed to CE research. The SLR shows AI techniques like supervised, unsupervised and reinforcement learning, fuzzy logic, image recognition and neural network being popularly adopted in CE. Fuzzy Logic, Decision Support System, Decision Tree and Random Forrest, Support Vector Machines and CNN has been repeatedly used. The SLR also discovered that, manufacturing process development, food supply chain and e-waste management need further attention. However, this study uses the documents only from the SCOPUS database. Other databases like Web of Science and Google Scholar can be explored in future work. We used only VOS Viewer software for our bibliometric analysis study. Other analysis packages like the R package, Bibexcel, Gephi can be considered in the future. The adoption of artificial intelligence techniques from different perspectives like industry, economy, system management, design strategies could also be investigated. The potentiality of integrating the circular economy with other emerging technologies like blockchain, big data and the internet of things should be explored.